\documentclass{styles/osa-article}
\journal{oe}
\articletype{Research Article}
\usepackage{amsmath}

\begin{document}

\title{Fibre polarization state compensation in entanglement-based quantum key 
distribution}

\author{Yicheng Shi,\authormark{1}  Hou Shun
  Poh,\authormark{1} Alexander Ling,\authormark{1,2} Christian 
Kurtsiefer,\authormark{1,2,*}}

\address{\authormark{1}Centre for Quantum Technologies, National University of 
Singapore, 3 Science Drive 2, Singapore, 117543\\
\authormark{2}Department of Physics, National University of Singapore, 2 
Science Drive 3, Singapore, 117542}
\email{\authormark{*}phyck@nus.edu.sg}

\begin{abstract}
Quantum Key Distribution (QKD) using polarisation encoding can be hard to 
implement over deployed telecom fibres because the routing geometry and the 
birefringence of the fibre link can alter the polarisation states of the 
propagating photons. These alterations cause a basis mismatch, leading to an 
increased Quantum Bit Error Rate (QBER). In this work we demonstrate a 
technique for dynamically compensating fibre-induced state alteration in a QKD 
system over deployed fibre. This compensation scheme includes a feedback loop 
that minimizes the QBER using a stochastic optimization algorithm.
\end{abstract}

(This version: \today)

\section{Introduction}
As first proposed in 1984, Quantum Key distribution enables two users to share an identical, random key that remains unknown to any third parties~\cite{BB84}. In theory, unconditionally secure communication can be established when QKD is used in conjunction with the one-time pad scheme~\cite{Vernam:1926,RevModPhys.81.1301}. A number of QKD protocols with proven security~\cite{Meyrs96,LoChau1999,ShorPreskill00} have been considered, which can be categorized as either "prepare-and-measure" schemes or entanglement based protocols. In practical implementations, qubits can be encoded into single photons (or approximations thereof) through their polarization or arrival times, and are transmitted between two parties either over free space to establish long distance links~\cite{Hughes2002,CK2002,PhysRevLett.98.010504,PhysRevLett.120.030501}, or through optical fibres for medium distance applications~\cite{Poppe:14,PhysRevX.8.021009,Dynes2019,Joshieaba0959}.

Encoding qubits into the polarisation of light has been widely adopted in many quantum information schemes, as different polarisation states can be easily prepared and measured for both weak optical pulses or single photons. Polarization encoded quibits are typically very robust against decoherence when propagating though free space or optically isotropic media. However, polarisation encoding faces a particular drawback as an optical fibre is not a pure loss channel for transmitting the polarisation states of photons. When propagating through the fibre, the state of polarisation (SOP) of a photon is altered due to the birefringence as well as the routing geometry of the fibre~\cite{Rashleigh:78}. In particular, fibre birefringence can be sensitive to changes in the ambient environment which makes this alteration somewhat random and time dependent~\cite{Xavier_2009}. This fibre-induced state alteration (or a rotation of a polarization when characterized as a point on the Poincar\'e sphere) causes basis mismatch, and eventually leads to an increased quantum bit error rate (QBER) in a QKD system, eventually preventing keys from being generated. Moreover, chromatic dispersion and polarisation mode dispersion of an optical fibre also degrades the timing correlation and degree of polarisation of the transmitted photons and further introduce errors to the system~\cite{RevModPhys.74.145,Hubel:07}.

While the dispersion effects of optical fibres can be mitigated with dispersion-shifted fibres or simply narrowing the optical bandwidth of the photons~\cite{Treiber_2009}, fibre-induced polarisation alteration needs to be actively monitored and compensated. This is usually achieved by placing a polarisation controller in the fibre link which is controlled with a feedback loop. The polarisation controller is set to implement a unitary transformation that inverts the polarization alteration of fibre. The resulting transformation of the entire channel is neutralized to the identity such that the polarization state of photons transmitted through the fiber remains unchanged. 

The optimal setting of the controller can be found by measuring the polarisation of two reference signals sent across the same fibre. This pair of reference signals needs to be prepared into two non-orthogonal polarization states, and the polarisation controller then is adjusted to reach a configuration where it restores the states of both reference signals at the output of the fibre. The reference signals can co-exist with the QKD photons in the same fibre via either time-division or wavelength-division multiplexing~\cite{Xavier:08, Chen_2009, Li:18}.
This type of compensation can operate at a high bandwidth at the cost of increasing hardware complexity, and is suitable for QKD systems with rapidly-oscillating environmental noise~\cite{Li:18}.

A different compensation scheme was proposed more recently that does not
require any reference light signals~\cite{Ding:17,Agnesi:20}. In this scheme,
one utilizes the number of erroneous bits in the revealed portion of the
sifted keys during error correction process, which has to be monitored in a
QKD protocol anyways to assess potential information leakage to an
eavesdropper. This error rate, which is an estimation of the system's QBER, is
used to generate an error signal for the polarisation controller. This
compensation simplifies the physical setup at the cost of a relatively low 
bandwidth of the feedback loop~\cite{Agnesi:20}.

In this work, we present a similar polarisation compensation technique,
but implement it in a polarisation-entanglement
based QKD system~\cite{Shi:2020}. Our technique uses a stack of liquid crystal 
variable retarders as polarisation controller and is optimized in a feedback 
loop using the estimated QBER as error signal. We also show that for 
polarization-entanglement based QKD, this technique exploits the rotational 
invariance of the distributed entangled state and only requires one of the two 
fibre links to be compensated. The compensation setup is implemented in a QKD 
system over a deployed telecom fibre link and achieves optimal compensation in 
under 20\,minutes. This technique requires minimal hardware overhead and is 
suitable for fiber-based QKD systems with slowly drifting environmental noise.

\section{Experimental setup}
\begin{figure}
  \centering
  \includegraphics[width=0.95\linewidth]{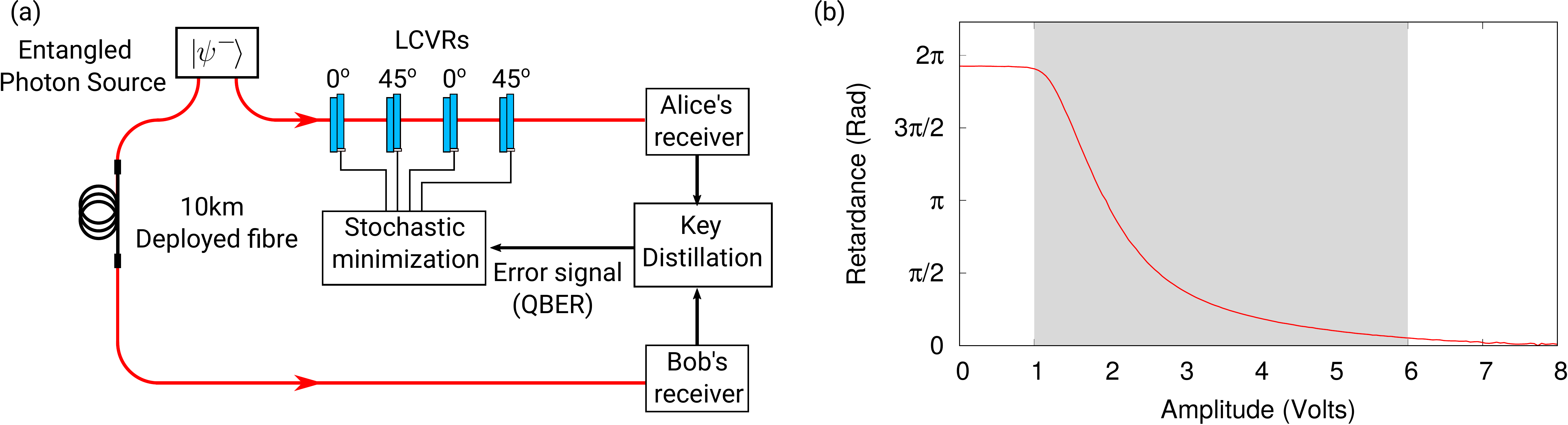}
\caption{(a) Experimental setup of a polarisation entanglement QKD system 
implemented over 10\,km of deployed fibre link. The polarization compensation setup consists of 4 liquid crystal variable retarders placed before Alice's receiver. (b) LCVR retardance versus applied voltage amplitude of 2\,kHz square wave at 1310\,nm.}
\label{fig:pol_lock_setup}
\end{figure}

A simplified diagram of our QKD setup with polarisation compensation is shown
in Fig.~\ref{fig:pol_lock_setup} (a). An entangled photon pair source prepares
photons pairs in a state $|\psi^-\rangle=\frac{1}{\sqrt{2}}(|H_A V_B\rangle - |V_A H_B \rangle)$. The photons are generated via Spontaneous Parametric Down-Conversion (SPDC), which converts pump light at 658\,nm to a signal and idler
photon at around 1310\,nm. The signal and idler photons are sent to two receivers, Alice and Bob,
respectively.  The signal photons are transmitted through a deployed telecom
fiber to Bob, while Alice receive the idler photons locally via a short
patchcord. The two receivers follow the BBM92 
protocol\cite{PhysRevLett.68.557} and randomly measure the polarisation of each 
photon in one of two bases: horizontal/vertical and diagonal/anti-diagonal. The 
basis is randomly chosen through a non-polarizing beam 
splitter\cite{Rarity:1994}, and exchanged between the two receivers via a 
classical channel during the key sifting procedure. Error correction is applied, 
which also allows to estimate an eavesdropper's potential knowledge of the key, 
and corresponding privacy amplification is applied to generate the final keys.

\begin{figure}
  \centering
  \includegraphics[width=0.95\linewidth]{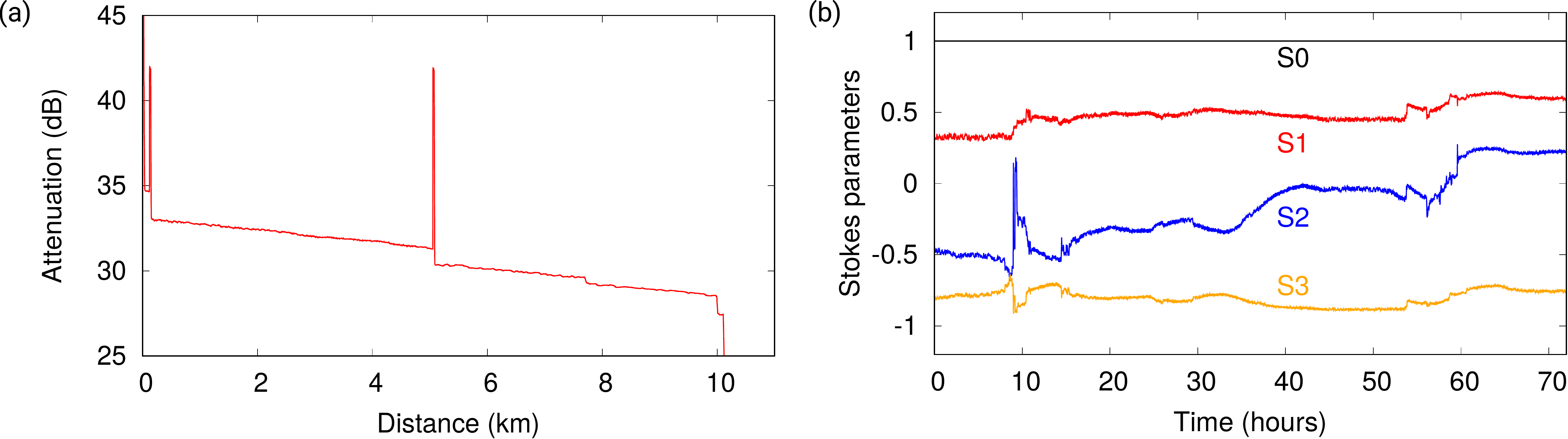}
\caption{(a) OTDR trace of the 10\,km deployed fibre. (b) Stokes parameters of 
polarisation state at the fibre output logged over 3 days showing drifts on a time scale of days.}
\label{fig:fibre-combined}
\end{figure}

As shown in Fig.~\ref{fig:fibre-combined} (a), the telecom fibre is about
10\,km long with approximately 7\,dB of optical attenuation. To simplify
experimental procedures, the fibre is deployed underground in a loop
configuration with both ends connected to the lab. The stability of the
deployed fibre is tested by sending in light with fixed polarisation and
monitoring the output state with a
polarimeter~\cite{doi:10.1080/09500340600674242}.
Figure~\ref{fig:fibre-combined} (b) shows a  72-hour measurement; the Stokes
parameters of the output state show only a slow drift with occasional
jumps. The change of the polarization state due to environmental influence
appears only to take place on a time scale of several minutes. 

To compensate for this slow drift of polarization state transmitted over fiber, a
polarization controller based on Liquid Crystal Variable Retarders (LCVRs) is
adequate. The reaction time of the LCVRs was measured to be about 5\,ms, which is sufficiently fast to compensate
the polarsation drifts we encounter. Moreover, the LCVRs include no
macroscopically moving parts and offer a high transparency at telecom
wavelengths (>95\%). A set of four LCVRs is placed before Alice's receiver to
serve as the polarisation controller. Each LCVR can provide a
voltage-controlled retardance from 0 to about $\frac{3}{2}\pi$ at 1310\,nm
(see Fig.~\ref{fig:pol_lock_setup} (b)). The LCVRs' optical axes are oriented
at $0^\circ$, $45^\circ$, $0^\circ$, and $45^\circ$ to allow for sufficiently
independent polarization transformations (see Fig.~\ref{fig:pol_lock_setup} (a)). While an arbitrary polarization transfer is completely described by a rotation direction and angle in the Poincar\'e
sphere, and thus 3 degrees of freedom should be sufficient to encode any
transformation required by the compensator, we chose four polarization
retarders to ensure that there is a continuous evolution of the control
parameters within their limited range, and that a gimbal lock situation is avoided.
In this way, any continuously varying unitary transformation between any arbitrary pairs of input
and output states can be implemented. 

\section{Polarization compensation for entangled states}
While QKD implementations based on "prepare-and-measure" protocols only
require a single fibre linking the sender and receiver, an implementation
based on entangled photon pairs needs two fibers to distribute photons to
both receivers. In this case, both fibers will alter the polarization states
of propagating photons. However, it is sufficient to only use a single
polarization compensator in one of the fibers, as the polarization of both
photons are correlated.

To see this, consider a source that generates photon pairs in a
rotationally invariant Singlet polarization state
$|\psi^- \rangle=\frac{1}{\sqrt{2}}(|H_A V_B\rangle - |V_A
H_B\rangle)$. Photons A and B undergo different fibre-induced polarisation
rotations $\hat{R}_A$ and $\hat{R}_B$. The resulting
photon pair state is
$(\hat{R}_A \otimes \hat{R}_B) |\psi^- \rangle$.
A polarisation controller acting on photon A can
be set to perform a transformation $\hat{T}_A$ such that $\hat{T}_A \hat{R}_A=\hat{R}_B$.
The resulting state
\begin{equation*}
(\hat{T}_A \hat{R}_A \otimes \hat{R}_B) |\psi^-\rangle =(\hat{R}_B \otimes
\hat{R}_B) |\psi^-\rangle = |\psi^-\rangle
\end{equation*}
is again the singlet state $|\psi^-\rangle$ due to its rotational invariance.
Thus, a single polarization compensation operation on one side is
sufficient to remove the state-changing actions of the fiber rotations
$\hat{R}_A$ and $\hat{R}_B$ on both transmission channels.

\begin{figure}
  \centering
  \includegraphics[width=0.95\linewidth]{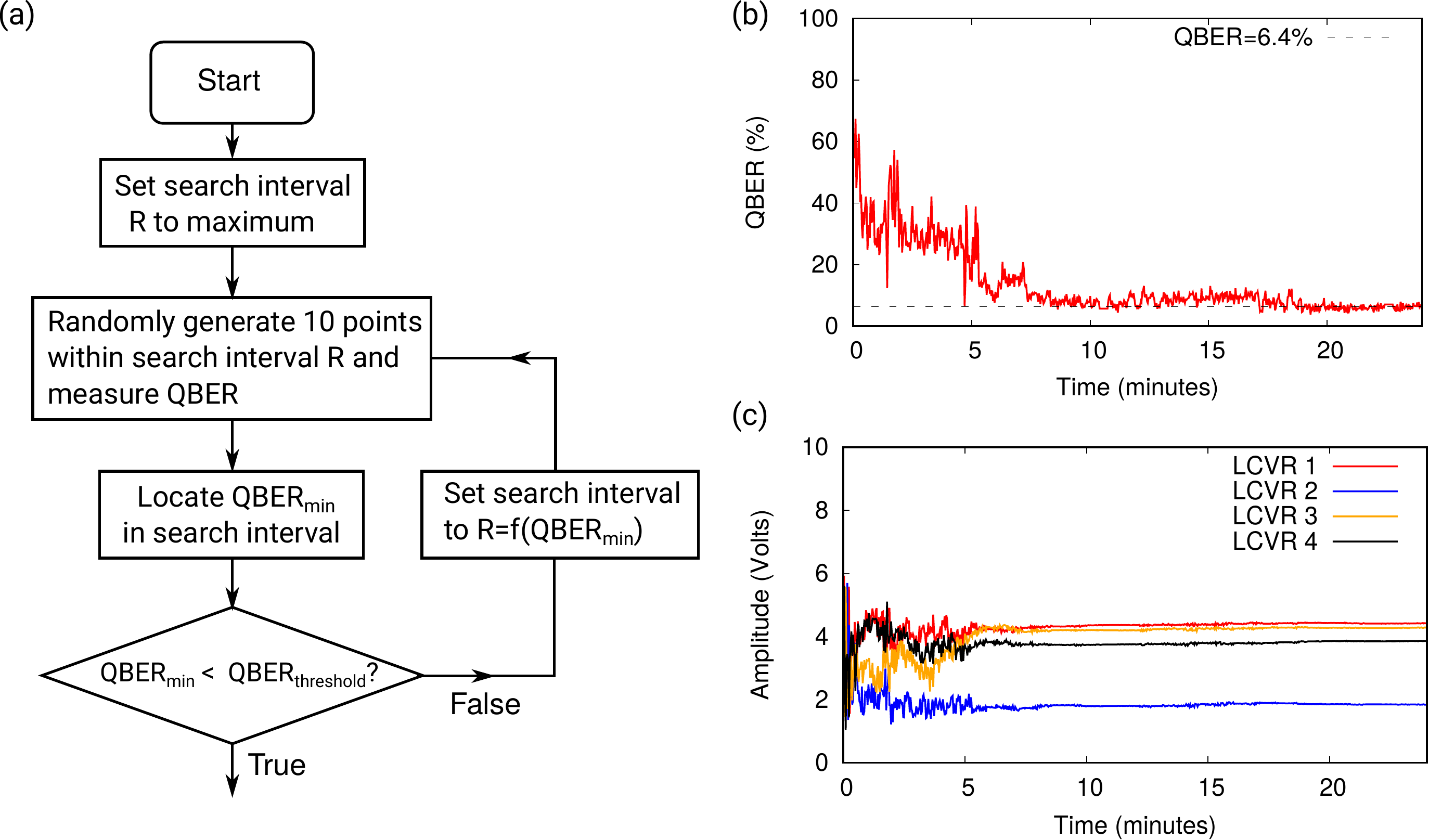}
\caption{(a) Flow chart of the stochastic search algorithm. (b) System QBER recorded during the stochastic search. (b) Applied voltage amplitudes for the LCVRs during stochastic search.}
\label{fig:search_algo}
\end{figure}

\section{QBER minimization with stochastic method}
With the setup shown in the previous section, the control loop for the
polarisation compensation can be considered as an optimization problem. The
goal of this optimization is to find the minimum of the estimated QBER of the
QKD system. This is now considered as a function of four variables,
$\text{QBER}=f(V_1,V_2,V_3,V_4)$, namely the control voltages $V_{1\ldots4}$
of the LCVRs. While this minimization problem can be solved using
gradient-descend algorithms in principle, we adopted a different approach in
this work due to practical considerations.

Firstly, it is impractical to obtain an accurate expression of the estimated
QBER as a function of the control voltages as the response curve of a LCVR
varies from unit to unit. Secondly, the estimated QBER cannot be measured with
a very high accuracy due to the limitation of finite sample
sizes. These limitations make it difficult to compute the gradients of
$f(V_1,V_2,V_3,V_4)$ from measurements, and a gradient-descend algorithm
cannot be efficiently implemented. Instead, we use a stochastic search algorithmdepicted in Fig.~\ref{fig:search_algo} (a).

The algorithm conducts a random search within a finite 4-dimensional parameter
space $(V_1, V_2, V_3, V_4)$. Each
control voltage takes a value between 1\,V and 6\,V which corresponds to
retardation from 0 to about $\frac{3}{2}\pi$ at 1310\,nm. The search algorithm
randomly picks a set of sample points in the entire parameter space and
measures the QBER for each point. The point with the smallest QBER in the
set will be chosen as the center of a next search iteration, which will be
conducted with the same number of points within a parameter hypercube of
smaller size $R$.
The size $R$
decreases with decreasing minimal QBER obtained in each iteration. As the
algorithm proceeds, the center point 
of the search will gradually approach the minimum in the entire space.

During QKD operation, the two receivers registered a coincidence rate of about
670\,s$^{-1}$ and a sifted key rate of 340\,s$^{-1}$ after basis
reconciliation. To reduce Poissonian noise, the
system QBER is evaluated from sifted keys accumulated over every 2
seconds. A typical starting condition before polarization compensation leads
to a QBER of
58\,$\pm$\,2.6\%, where the uncertainty is infered from the Poissonian
counting statistics. With this initial QBER,
the stochastic search begins its first iteration with a set of 10 points.
The reduction of the search range $R$ in the parameter space in the
iteration is accomplished with an ad-hoc chosen function $R=A\times
(\text{QBER}_{\text{min}}-\text{QBER}_{\text{threshold}})^B$, where
QBER$_\text{min}$ is the minimal QBER in any given iteration. The coefficients
$A$ and $B$ set the rate at which the search algorithm converges to the global
minimum, while the offset $\text{QBER}_{\text{threshold}}$ sets a lower bound
of the QBER given by other elements than the optical fiber in the QKD
system. The last choice assures that the parameter space is still probed in
a reasonable neighborhood of the global QBER minimum. Continuously operating
this algorithm allows to follow a drift of this minimum location in the
parameter space over time in a control-loop-like fashion. We found that in our
system, a choice of $A=6.5\,\text{Volts}$, $B=2$, and
$\text{QBER}_{\text{threshold}}=4\%$ worked well.

Fig.~\ref{fig:search_algo} (b) shows the performance of our polarisation
compensation technique in an exemplary single run. The stochastic search
algorithm reduces the system QBER from its initial value of 58\,$\pm$\,2.6\%
to about 7\,$\pm$\,0.7\% after about 10\,minutes (about 30 iterations of
search). We then observed a small increase of QBER by about 3\%, possibly due
to a disturbance to the fiber, but the algorithm  eventually lowers the QBER
down to 6.4\,$\pm$\,0.7\%. The corresponding control voltages of the LCVRs
during the search process are shown in Fig.~\ref{fig:search_algo} (c). They
converge to stable values as the QBER approaches its minimum given by other
system consrtraints.

\section{Conclusion}
We demonstrated polarisation compensation in an entanglement-based QKD system 
over a deployed telecom fibre. This technique, which utilizes the estimated 
QBER as the error signal for a feedback control loop, does not require any 
reference light sources or extra detectors in the setup. We show that by 
exploiting the rotational invariance property of the Bell $|\Psi^-\rangle$ 
state, one only needs to apply compensation of one of the fibre links in an 
entanglement QKD system. The control loop of the polarisation compensation runs 
a stochastic search algorithm that actively minimizes the estimated QBER and is 
able to achieve optimal compensation in under 20\,minutes. 

While this technique is slower compared to methods based on reference signals
used to measure out the fiber transformation, it is very simple to implement
and requires minimal hardware overhead. The only hardware required is a
polarisation controller.
This technique is suitable for deployed fibre-QKD systems with slowly drifting
environmental polarization noise. The compensation process does not leak any
information through any channels, and therefore does not compromise the security of the QKD link.

\section*{Funding}
This  research  was  supported  by  the  National  Research Foundation, Prime 
Minister’s Office, Singapore under its CorporateLaboratory@University Scheme, 
National University of Singapore, and Singapore Telecommunications Ltd.


\bibliography{reference}
\end{document}